\documentclass[journal]{IEEEtran}
\IEEEoverridecommandlockouts
\usepackage{cite}
\usepackage{amsmath,amssymb,amsfonts}
\usepackage{algorithmic}
\usepackage{graphicx}
\usepackage{textcomp}
\usepackage{xcolor}
\def\BibTeX{{\rm B\kern-.05em{\sc i\kern-.025em b}\kern-.08em
    T\kern-.1667em\lower.7ex\hbox{E}\kern-.125emX}}
\begin{document}

\title{Design of a System for Analyzing Cell Mechanics}

 \author{
    \IEEEauthorblockN{Hasan Berkay Abdioglu\(^1\), Yagmur Isik\(^1\), Merve Sevgi\(^2\), Ufuk Gorkem Kirabali\(^1\), Yunus Emre Mert\(^1\),\\ Gulnihal Guldogan\(^3\), Selin Serdarli\(^4\), Tarik Taha Gulen\(^5\), Huseyin Uvet\(^1\)\\}
\vspace{4mm}
    \IEEEauthorblockA{\(^1\)Department of Mechatronics Engineering, Yildiz Technical University, Istanbul, Turkey\\}
    \IEEEauthorblockA{\(^2\)Department of Bioengineering, Yildiz Technical University, Istanbul, Turkey\\}
    \IEEEauthorblockA{\(^3\)Cakir School, Bursa, Turkey}
    \IEEEauthorblockA{\(^4\)Modern Innovation Lyceum, Baku, Azerbaijan\\}
    \IEEEauthorblockA{\(^5\)SEV American College, Istanbul, Turkey}
}

\maketitle

\begin{abstract}

Accurately measuring cell stiffness is challenging due to the invasiveness of traditional methods like atomic force microscopy (AFM) and optical stretching. We introduce a non-invasive off-axis system using holographic imaging and acoustic stimulation. This system features an off-axis Mach-Zehnder interferometer and bulk acoustic waves to capture cell mechanics. It employs high-resolution components to create detailed interferograms and allows continuous imaging of cell deformation. Unlike conventional techniques, our method provides high-throughput, label-free measurements while preserving cell integrity. Polyacrylamide beads are tested for high precision, highlighting the potential of the system in early cancer detection, disease monitoring, and mechanobiological research.

\end{abstract}


\section{Introduction}
Cell stiffness is a crucial mechanical property of biological cells, fundamental to understanding various cellular processes including migration, division, differentiation, and adhesion \cite{b1,b2,b3}. This property arises from the intricate interplay among components within the cell, such as the cytoskeleton, extracellular matrix, and membranous organelles. The field of mechanobiology, which explores these interactions and the cellular responses to mechanical stimuli, is essential for understanding both cellular physiology and pathology \cite{b4,b5,b6,b7}.

The deformation of cells is closely linked to their internal components like filaments and the extracellular matrix. Each component's deformation is influenced by its geometry, properties, microstructure, and external loads. By studying the mechanical properties of individual cells, we can gain significant insights into these complex processes. For instance, cell stiffness is integral to biological activities related to growth, mobility, division, and differentiation. The cytoskeleton, particularly actin filaments and intermediate filaments, significantly affects cell mechanics, and changes in these structures can directly alter cellular mechanical properties \cite{b8,b9,b10,b11,b12}.

Variations in cell stiffness are indicative of different cellular conditions and diseases. For example, cancer cells typically show reduced stiffness due to fewer and irregular actin filaments, facilitating their migration and invasiveness. Research in ovarian and breast cancers has demonstrated that decreased cell stiffness correlates with cytoskeletal remodeling, suggesting that cell stiffness could serve as a biomarker for metastatic potential \cite{b13,b14,b15,b16,b17}.

\section{Proposed System}
Considering these challenges, there is a critical demand for advanced techniques that are non-invasive, rapid, and capable of high-throughput screening. Such methods would greatly enhance both clinical diagnostics and mechanobiological research by providing more accurate and efficient assessments of cell stiffness. This advancement would not only deepen our understanding of cellular mechanics but also aid in the creation of new diagnostic tools and therapeutic strategies, especially in cancer research\cite{b17,b18}.

Our research introduces an innovative cell stiffness measurement sensor based on an off-axis Mach-Zehnder interferometer. This approach addresses the shortcomings of current methods by allowing for continuous imaging and direct observation of the mechanical effects of acoustic vibrations. The design of the chip was meticulously evaluated to minimize deviations in acoustic stimulation, optimize the chip's metrics and transducer placement, and establish robust procedures for image acquisition.

\begin{figure}
    \centering
    \includegraphics[width=1\linewidth]{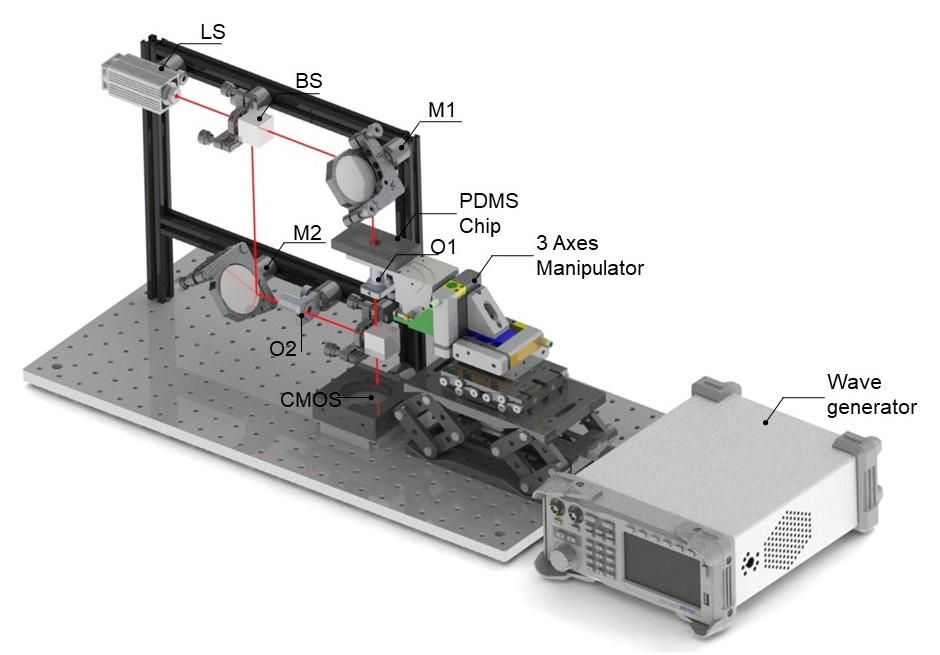}
    \caption{Schematic of the Off-Axis Mach-Zehnder Interferometer Setup. (LS, laser source; BS, beam splitter; M1, M2, mirror; O1, O2, objective lens ). The laser beam is split into two paths: one beam passes through the object (such as a cell or tissue), while the other serves as the reference beam. These beams are then recombined to form an interferogram, which is captured by a high-speed CMOS camera. This setup enables continuous imaging and direct observation of mechanical effects induced by acoustic vibrations, allowing for precise measurement of cell stiffness.}
    \label{fig:enter-label}
\end{figure}

\section{Methods and Materials}
\subsection{Production of Transducer-Integrated PDMS Chip}
To measure cell stiffness, we utilized transducer-integrated Poly(dimethylsiloxane) (PDMS) chips. The design of these chips was created using SolidWorks, followed by the fabrication of molds through aluminum milling. The microfluidic chip, measuring 24 x 60 x 10 mm, was produced from PDMS material using soft lithography. A mixture of silicone elastomer and curing agent at a 10:1 ratio was poured into the mold. To remove bubbles, the mixture was placed in a desiccator for 45 minutes. Afterward, the PDMS was cured by baking at 65°C for 3 hours. The cured PDMS was then carefully extracted from the mold using a scalpel. Finally, the fabricated PDMS structure was bonded with the transducer using an oxygen plasma machine at 600 mTorr pressure for 50 seconds.

\subsection{Experimental Setup}
The experimental setup was divided into two distinct subsystems: one for imaging, which captures three-dimensional vibration data through a holographic setup, and the other incorporating the chip with an integrated transducer to generate cell vibrations. A 3-axis stage (PI M-126 Precision Microtranslation Stage) was employed to ensure complete imaging of cells across the chip and precise focus adjustment.

The holographic setup is based on an off-axis Mach-Zehnder Interferometer. By varying the angle of the reference beam, phase extraction can be rapidly achieved. The interferometric setup includes a 671 nm wavelength, 200 mW laser, two beam splitters, two mirrors, and two 20X 0.40 NA objectives. Interferograms are recorded at high speed (200 fps) with a Basler boost boA1936-400cm CMOS camera, featuring a 4.5 
\(\mu\)mpixel size. The laser beam's power is attenuated using a neutral density filter at the laser output to ensure accurate imaging of objects.

\subsection{Processing of Images}
\subsubsection{Holographic Reconstruction}
Three-dimensional holographic images are obtained by processing off-axis hologram recordings using a Mach-Zehnder interferometer. As illustrated in Figure 1, the laser beam is initially split into two. One beam passes through an object, such as a cell or tissue, while the other beam serves as the reference beam and does not interact with the object. These beams are then recombined through a beam splitter to form an interferogram.

The intensity of the images captured in the microscope setup based on the Mach-Zehnder interferometer can be expressed as:

\[I = I_1^2 + I_2^2 + I_1 I_2^* + I_1 I_2
\]

Where \(I_1\) is the intensity of the reference beam, and \(I_2\) is the intensity of the object beam. The first two terms of the equation are considered constant, known as the DC term. The third and fourth terms are respectively named the -1st order and 1st order terms. The information used is carried by the fourth term of the equation, i.e., the 1st order term. The fourth term has both amplitude and phase information of the object that is illuminated by the object beam. The phase of the object is related to depth information.

\subsubsection{Reconstruction of Stiffness Maps}

To measure cell stiffness using an acoustic-holographic setup, we integrate acoustic and optical systems. Surface acoustic waves in the cell medium are generated using a lead zirconate titanate (PZT) transducer. These waves induce mechanical stimuli, and the cells' responses are captured using Digital Holographic Microscopy (DHM) with an off-axis Mach-Zehnder interferometer and a high-speed CMOS camera.

As the acoustic waves propagate, they cause mechanical changes in the cell membrane, which are recorded as interferograms by DHM. Interferograms result from the interference of light waves between the reference and object beams, highlighting changes in cell thickness and morphology due to the acoustic waves.

Each captured frame corresponds to specific points in the cell's vibration cycle. By sequentially processing these frames, we generate thickness maps and observe membrane vibrations, visually depicting the morphological changes in cells.

Cell stiffness is calculated using the Hertz elasticity model, which assumes cells behave as isotropic, linear elastic solids in an infinite half-space. The mechanical impact of acoustic pressure on the cell membrane is assessed by calculating the pixel-wise change in cell thickness, thus linking acoustic pressure to membrane vibration and determining cell stiffness.

The displacement \(d\) at the moment of vibration obtained from holographic images and the displacement at rest \(h_0\) can be used to calculate the displacement:
\[d = h - h_0
\]

The cell strain \(\epsilon\) is then calculated as:

\[\epsilon = \frac{d}{h_0}
\]

The expression for the elastic modulus 
\(\)
\textit{E} (stiffness) for each pixel 

\textit{j} is defined as the ratio of the stress exerted on the object 
\(\Pi\) to the corresponding strain 
\(\epsilon\), given by the equation:

\[E_j = \frac{\Pi_j}{\epsilon}
\]

\section{Results}

In our study, we developed a high-precision sensor to measure cell stiffness using acoustic stimulation and off-axis holography. A standardized measurement protocol was established, with the vibration frequency set at 10 Hz and the camera capture speed at 200 fps. Polyacrylamide beads (\(\leq\)45µm) were used for testing due to their similar stiffness to cells and were affixed to the chip with CELL-TAK to simulate adherent cells.

The average stress on the bead surface was 75 Pa, resulting in an average stiffness of 1828 Pa, which closely aligns with previous measurements. For relative stiffness comparisons, consistent imaging squares are recommended, while varied squares are sufficient for distinct stiffness evaluations.

\section{Discussion}

The system designed in this study offers significant improvements over traditional cell stiffness measurement techniques such as atomic force microscopy, optical tweezers, and micropipette aspiration. It is non-invasive, preserving cell integrity and viability, and label-free, avoiding artifacts and allowing cells to be studied in their natural state. The system enables rapid data acquisition for real-time analysis and high-throughput screening, crucial for both research and clinical diagnostics. High-resolution holographic imaging provides detailed visualization of cell morphology and mechanical responses, capturing subtle changes missed by other techniques.

The accuracy and efficiency of this sensor in measuring cell stiffness have important implications for cancer diagnostics. It allows differentiation between cancerous and healthy cells, monitoring disease progression, and assessing treatment efficacy. It also aids in mechanobiology research, offering insight into cell stiffness in various processes, which could lead to new therapeutic strategies.

Future research should focus on miniaturizing the system for clinical integration, validating its efficacy with diverse cell types and disease models, and exploring its integration with other diagnostic tools for comprehensive diagnostic platforms.

\section{Conclusion}

This study introduces a system for analyzing cell mechanics, advancing the fields of cell mechanobiology and cancer diagnostics. Utilizing real-time holographic reconstruction and acousto-holographic measurement, it addresses the limitations of traditional methods.

Key achievements include high precision and repeatability in stiffness measurement, which are crucial for diagnostics and research. Its non-invasive nature preserves cell integrity, making it ideal for delicate samples. Rapid, label-free measurements enable real-time monitoring and high-throughput screening, critical in clinical settings. Off-axis Mach-Zehnder interferometry provides high-resolution imaging, revealing detailed cell morphology and mechanical changes.

The sensor's capability to differentiate cells based on stiffness holds promise for enhancing cancer diagnostics by enabling early detection and monitoring disease progression. In mechanobiology, it offers insights into cellular mechanics, potentially leading to new therapies.

Future work includes miniaturizing the sensor for clinical integration, validating its use in various types of cells and diseases, and integrating it with other diagnostic tools to create comprehensive platforms.

In summary, this sensor represents a transformative approach in biomechanical measurements, promising precise assessments of cell stiffness with broad implications for biomedical research and diagnostics, potentially significantly advancing healthcare.

\vspace{12pt}

\end{document}